\RequirePackage[l2tabu,orthodox]{nag}
\documentclass[11pt]{article}
\pdfoutput=1
\usepackage{times}
\usepackage{amsmath}
\usepackage{amssymb}
\usepackage{latexsym}
\usepackage{graphicx}
\usepackage{subfigure} 
\usepackage{url}
\usepackage[stretch=10]{microtype}
\usepackage{todonotes}
\usepackage{algorithm}
\usepackage{algorithmic}

\begin{document}

\title{Algorithm 950: Ncpol2sdpa---Sparse Semidefinite Programming Relaxations for Polynomial Optimization Problems of Noncommuting Variables}
\author{Peter Wittek\\
 \small{University of Bor\aa{}s}
}
\date{}
\maketitle
\begin{abstract}
A hierarchy of semidefinite programming (SDP) relaxations approximates the global optimum of polynomial optimization problems of noncommuting variables. Generating the relaxation, however, is a computationally demanding task, and only problems of commuting variables have efficient generators. We develop an implementation for problems of noncommuting problems that creates the relaxation to be solved by SDPA -- a high-performance solver that runs in a distributed environment. We further exploit the inherent sparsity of optimization problems in quantum physics to reduce the complexity of the resulting relaxations. Constrained problems with a relaxation of order two may contain up to a hundred variables. The implementation is available in Python. The tool helps solve problems such as finding the ground state energy or testing quantum correlations.
\end{abstract}

\section{Introduction}
In polynomial optimization, we are interested in finding the global minimum $p^{\star}$ of a polynomial $p(x)$:
\begin{eqnarray}
p^{\star}& = & \min_{x} p(x),\\
\mathrm{s.t.} & & g_{i}(x)\geq 0, \quad i=1,\ldots,m,
\label{commutative}
\end{eqnarray}
where the variables $x=(x_{1},\ldots,x_{n})\in\mathbb{R}^{n}$, and $g_{i}(x)\geq 0, i=1,\ldots,m$ are polynomials. For this case, Lasserre introduced a sequence of semidefinite program (SDP) relaxations that converge to the global optimum~\cite{lasserre2001global}. SDPs are a widely studied area of convex optimization and related problems appear in other areas of science~\cite{vandenberghe1996semidefinite}. While the sequence of SDPs is infinite and grows polynomially in size, convergence may happen fast.

The key idea of the method is to decompose the polynomial $p(x)$ in the basis of monomials $w(x)$ as $p(x)=\sum_{w}p_{w}w(x)$. By treating every monomial as an independent variable $y_{w}$, the task is to minimize the linear combination $\sum_{w}p_{w}y_{w}$. Since each $y_{w}$ corresponds to a monomial, they are not independent, and they should satisfy non-convex polynomial constraints. The non-convex constraints are relaxed by requiring the positivity of a moment matrix, whose entries are indexed in the basis of monomials and are given by $M(y)(v,w)=y_{vw}$. The constraints $g_{i}(x)\geq 0$ that are relaxed by imposing the positivity of a set of localizing matrices $M(g_{i}y)(v,w)=\sum_{u}g_{u}y_{uvw}$. We replace the optimization problem (\ref{commutative}) by the following SDP:
\begin{eqnarray}
\min_{y}      &   \sum_{w}p_{w}y_{w}\label{relaxed} \\
\mathrm{s.t.} &   M(y)\succeq 0, \nonumber \\
              &  M(g_{i}y)\succeq 0, i=1,\ldots,r,\nonumber
\end{eqnarray}
where $\succeq 0$ stands for expressing the positive definiteness of the left hand side.

The moment and localizing matrices are infinite, but they can be truncated at any finite size $k$. Increasing the order of the truncation yields higher quality approximation of the original problem (\ref{commutative}).

The same idea of using a hierarchy of SDPs applies to the solution of polynomials of noncommuting variables~\cite{pironio2010convergent,navascues2012sdp}. The sequence converges provided some conditions on the operators, and in some cases it is possible to conclude that the minimum has been reached after performing only a finite number of tests~\cite{navascues2008convergent}.

To solve the SDP, we must convert the noncommuting polynomial optimization problem to an SDP of a given relaxation order. This hinges on symbolic manipulations to extract the numerical problem at hand. The conversion itself is a costly operation. Thus, we developed Ncpol2sdpa, an efficient library to perform the conversion that also considers the sparsity of the relaxations of polynomial optimization problems.

\section{Relaxations of polynomial optimization problems of noncommuting variables}\label{generate}
We follow \cite{navascues2012sdp} to introduce the hierarchy of SDP relaxations more formally. The formulation of a polynomial optimization problem of noncommuting variables is similar to the commuting case in Eq.~(\ref{commutative}). Consider the set of $2n$ letters $(x_{1} , x_{2} , \ldots, x_{n} , x_{1}^{*} , x_{2}^{*} , \ldots, x_{n}^{*})$. We form a word $w$ of length $k$ by joining $k$ such letters in a particular order. An empty word is denoted by 1 -- its length is zero by definition. The set of all words of a length up to $k$ is denoted by $W_{k}$, whereas $W$ denotes the set of all words. 

We define involution on a word $w = w_1 w_2\ldots w_k$ as $w^{*} = w_k^{*} w_{k-1}^{*} \ldots w_1^{*} $. For individual letters, the involution is defined as $(x_i)^{*} = x_i^*$ and $(x_i^{*})^{*} = x_i$ .

With this notation, let us consider the algebra $\mathbb{K}[x,x^{*}]$ of polynomials of $2n$ noncommuting variables $(x_{1} , x_{2} , \ldots, x_{n} , x_{1}^{*} , x_{2}^{*} , \ldots, x_{n}^{*})$ with coefficients in $\mathbb{K}$, where $\mathbb{K}\in\{\mathbb{R},\mathbb{C}\}$. An element $p$ in this algebra is the following linear combination:
\begin{equation}
\sum_{w}p_ww,
\end{equation}
where $w\in W$, and the summation is finite. Involution in this sense is essentially a conjugate transpose. The degree of a polynomial $p$ is the length of the longest word. A polynomial is said to be Hermitian if $p^{*}=p$. Words in this sense correspond to monomials.

Furthermore, let $\mathcal{B}(H)$ denote the set of bounded operators on a Hilbert space $H$ defined on the field $\mathbb{K}$. We consider a set of operators $X = (X_1 ,\ldots, X_n )$ from $\mathcal{B}(H)$. Given a polynomial $p \in \mathbb{K}[x, x^∗]$, we define the operator $p(X) \in \mathcal{B}(H)$ by substituting every variable $x_{i}$ by the operator $X_{i}$, and interpreting involution as the adjoint operator on $H$. If the polynomial is Hermitian, its corresponding operator $p(X)$ will also be Hermitian, thus $\langle\phi,p(X)\phi\rangle$ will be real for every vector $\phi$ in $H$.

Let $p$ and $g_{i}$ $(i = 1, \ldots, m)$ be Hermitian polynomials in $\mathbb{K}[x, x^{∗}]$. We write the optimization problem as follows:
\begin{eqnarray}
  p^{\star} & = & \inf_{X, \phi} \langle\phi, p(X)\phi\rangle \\
  \mathrm{s.t.} 
  & & \|\phi\| = 1, \nonumber\\
  & & g_{i}(X)\succeq 0, \quad i=1,\ldots,m.\nonumber
  \label{noncommuting}
\end{eqnarray}
Optimization is over all Hilbert spaces $H$, all sets of operators $X = (X_{1},\ldots, X_{n})$ in $B(H)$, and all normalized vectors $\phi$ in $H$.

To define the SDP relaxation, let $y = (y_{w})_{|w|\leq{}d} \in \mathbb{K}^{|W_{d}|}$ be a sequence of real or complex numbers indexed in $W_{d}$ . That is, we make a correspondence between each word $w$ in $W_{d}$ and a number $y_{w} \in \mathbb{K}$. We define a linear mapping $L_{y}: \mathbb{K}[x, x^{*}]_{d} \mapsto \mathbb{K}$ as
\begin{equation}
p \mapsto L_{y}(p) = \sum_{w\leq d} p_{w} y_{w}
\label{mapping}
\end{equation}
By this mapping, we match the variables $X$ in the Hilbert space $H$ with numbers in $\mathbb{K}$.

Given a sequence $y = (y_{w})_{|w|\leq{}2d}$ indexed in $W_{2d}$ , we define the moment matrix $M_{d}(y)$ of order $d$ as a matrix with rows and columns indexed in $W_{d}$ and whose entry $(v, w)$ is given by
\begin{equation}
M_{d}(y)(v, w) = L_{y} (v^{*} w) = y_{v^{*}w}.
\end{equation} 

If $g = \sum_{|u|\leq k}g_{u} u$ is a polynomial of degree $k$ and $y = (y_{w} )_{|w|\leq 2d+k}$ is a sequence indexed in $W_{2d+k}$ , we define the localizing matrix $M_{d}(gy)$ as the matrix with rows and columns indexed
in $W_{d}$, and whose entry $(v, w)$ is defined as
\begin{equation}
M_{d}(gy)(v, w)=L_{y}(v^{*}gw) = \sum_{|u|\leq{}k}g_{u}y_{v^{*}uw}.
\label{localizing}
\end{equation}

We define the SDP relaxation of Eq.~(\ref{noncommuting}) as follows. Given a relaxation order $d$, where $2d\geq  \max\{\deg(p), \max_{i}\deg(g_{i})\}$, the relaxation is given by
\begin{eqnarray}
\label{ncrelaxation}
  p^{d}         & = & \min_{y} \sum_{w}p_{w}y_{w}\\
  \mathrm{s.t.} &   &  y_{1} = 1 \nonumber \\
                &   & M_{d}(y)\succeq 0 \nonumber \\
                &   & M_{d-d_{i}}(g_{i}y)\succeq 0 \quad i=1,\ldots,m, \nonumber
\end{eqnarray}
where $d_{i}=\lceil\deg(g_{i})/2\rceil$, and the optimization is over $y=(y_{w})_{|w|\leq 2d}\in \mathbb{K}^{|W_{2d}|}$. The $y_1 = 1$ constraint comes from $y_1 = \langle\phi,\phi\rangle = \|\phi\| = 1$, where $\langle.,.\rangle$ stands for the inner product in $H$.

For increasing orders of $d$, the SDPs define a growing hierarchy, whose solution converges to the optimum of Eq.~(\ref{noncommuting})~\cite{pironio2010convergent}. It is worth noting that in some cases, such as in the ground-state energy problem of bosonic systems, the hierarchy already converges at first-order relaxations~\cite{navascues2013paradox}.

Starting from a problem defined in the form of Eq.~(\ref{noncommuting}) with noncommuting variables, our goal is to derive the relaxation in Eq.~(\ref{ncrelaxation}), and format the relaxation to fit a solver. The solver expects the following semidefinite programme:
\begin{eqnarray}
\label{sdpdefinition}
\min_{x} & & \sum_{l=1}^{n}c_{l}x_{k} \\
\mathrm{s.t.} & & \sum_{l=1}^{n}F_{l}x_{l}-F_{0}\succeq 0, \nonumber
\end{eqnarray}
where $F_{l}$ ($l=0,1,\ldots,n$) are the symmetric matrices called constraint matrices. A variable in the SDP formulation corresponds to a relaxation variable, hence the number of SDP variables is equivalent to the number of entries in the moment matrix $M(y)$. The size of the constraint matrices depends on the number of variables, the number of constraints, and the order of the relaxation.

\section{Algorithm description}\label{algorithm}
The algorithm can be broken down in five steps:
\begin{description}
\item[Step 1] Definition of polynomial optimization problem of noncommuting variables.
\item[Step 2] Generating the moment matrix.
\item[Step 3] Translating the objective function.
\item[Step 4] Generating the localizing matrices.
\item[Step 5] Exporting the SDP relaxation.
\end{description}

The most complex step is generating the moment matrix. This is also the computationally most demanding step where resource optimization is crucial. Obtaining the objective function and the inequalities are comparably easier.

The definition of the problem is requested from the user. High-level noncommuting operations are allowed to express the problem intuitively. The rest of the algorithm processes the problem definition, at the relaxation order required.

\begin{algorithm}
\caption{Generating the moment matrix}
\begin{algorithmic}
 \REQUIRE Noncommuting variables, order of relaxation.
 \ENSURE Moment matrix, monomial dictionary.
 \STATE Calculate monomials in $W_{d}$ for given order $d$.
 \FOR{$v\in W_{d}$}
   \FOR{$w\in W_{d}$}
     \STATE Get entry label $v^{*}w$ for $M_{d}(y)(v, w)$
     \STATE Call fast substitutions for $v^{*}w$
     \IF{monomial $v^{*}w$ appeared before}
       \STATE Find corresponding variable $L_{y}(v^{*}w)$.
       \STATE Insert value in moment matrix at the current location.
      \ELSE
        \STATE Create new variable $L_{y}(v^{*}w)$.
        \STATE Insert value in moment matrix at the current location.
     \ENDIF
   \ENDFOR
 \ENDFOR
\end{algorithmic}
\label{alg:momentmatrix}
\end{algorithm}

When generating the moment matrix (Algorithm~\ref{alg:momentmatrix}), we start with computing the moments in the first row, with elements $L_{y}(w)_{|w|\leq d}$. The outer product with the adjoint of this row will define the moment matrix, or, more precisely, the upper triangular part of the moment matrix.

Calculating the moment is one of the points where noncommuting problems differ from commuting ones. The implementation relies on external noncommuting libraries to get those moments. Although replicating the limited number of operations needed to get the moments would be easy, we reduce code complexity by using the symbolic operations of the supporting libraries.

The symbolic operations also include substitutions, but we use them for a very specific purpose, which forced us to write our own substitution routines. We return to this issue in Section~\ref{additional}.

\phantomsection
\label{par:numberofmonomials}
The size of this matrix depends on the number of monomials in the basis, which in turn depends on the number of words in the set $W_{d}$ of words up to degree $d$, where $d$ is the order of the relaxation. The number of words in $W_{d}$ is $|W_{d}| = \left((2n)^{d+1}-1\right)/(2n-1)$. If the noncommuting variables are Hermitian, the size of the alphabet reduces to half, with a corresponding change in the formula for $|W_{d}|$.

Unordered associative arrays are at the core of generating the moment matrix and handling moments. We index the associative arrays with the moments of noncommuting variables in the internal loop of generating the moment matrix -- in other words, we cache calls of the mapping function in Eq.~(\ref{mapping}). If we encounter a moment that was not seen before, we insert it in the associative array, noting its location within the moment matrix. We also push an entry to the moment matrix at the current location.

There will always be, however, moments that repeat -- the moment matrix has further symmetries other than being Hermitian. Therefore a search is performed at each iteration to find whether we have already seen a particular moment. The associative array has an average complexity of $O(1)$ for such searches. If we find a matching element, we insert that element in the moment matrix at the current location. By doing so, we save on the number of relaxation variables. If such replacement were not done, the symmetries would require the explicit encoding of equalities, which would in turn translate to pairs of inequalities, adding a tremendous number of additional localizing matrices to the problem. Given the constant complexity of searches, the overall average complexity is quadratic in the number of monomials in the basis of the relaxation.

\begin{algorithm}
\caption{Generating a localizing matrix}
\begin{algorithmic}
 \REQUIRE Polynomial, monomial dictionary, order of relaxation.
 \ENSURE Localizing matrix 
 \STATE Calculate maximum order $d$ of localizing matrix.
 \STATE Calculate monomials $W_{d}$ for given order $d$.
 \FOR{$v\in W_{d}$}
   \FOR{$w\in W_{d}$}
     \STATE Calculate entry $\sum_{|u|\leq{}k}g_{u}y_{v^{*}uw}$ for $M_{d}(gy)(v, w)$.
     \STATE Call fast substitutions.
     \STATE Match variables from monomial dictionary.
     \STATE Push corresponding entries to respective constant matrices $F_{k}$.
   \ENDFOR
 \ENDFOR
\end{algorithmic}
\label{alg:localizingmatrix}
\end{algorithm}
The objective function is a matter of looking up the members in the monomial dictionary. The computational time is negligible.

Generating the localizing matrices (Algoritm~\ref{alg:localizingmatrix}) also uses simple noncommutative operations to get the $M_{d}(gy)(v, w)$ elements of the matrices. Furthermore, we also use fast substitutions here (Section~\ref{additional}). Then the task becomes identical to the translation of the objective function, repeated for each $M_{d}(gy)(v, w)$ entry. We use searches in the associative arrays, which have an average complexity of $O(1)$. The localizing matrices are always smaller then the moment matrix, hence their generation is much faster.  Each inequality defines a $b\times{}b$ block, where $b$ is the number of monomials in the localizing matrices, $M(g_{i}y)$.

Equality constraints, if any, are transformed to pairs of inequality constraints, although this behaviour can be altered in special cases (see Section~\ref{additional}).

A key source of sparsity is the data structure storing the entries of the $F_{l}$ matrices in Eq.~(\ref{sdpdefinition}). An entry is structured as follows:
\begin{equation}
l\quad b\quad i\quad j\quad v,\nonumber
\end{equation}
where $l\in\{0,1,\ldots,n\}$ is the index of the variable,  $b\in\{1,2,\ldots,B\}$ is  the block index, $j\geq{}i\geq{}1$ are  the indices of the upper triangular entry of the constraint, and $v$ is the value of the constraint. Only nonzero entries are stored. This format is identical to the sparse input format expected by SDPA, also making exporting efficient, with minimal overhead. The last step of the algorithm thus is fast, its computational time is negligible.

\section{Implementation}\label{implementation}
We follow the general structure of other tools that format SDP problems to adapt to specific solvers, such as Yalmip in MATLAB~\cite{lofberg2004yalmip} and Picos in Python~\cite{sagnol2012picos}. Noncommuting variables only have a rudimentary support in MATLAB, and conversions are not scalable. Picos aims to be the Python-based equivalent of Yalmip, a generic wrapper to generate and format optimization problems. Unfortunately, problems generated with Picos cannot efficiently add polynomial elements to the block matrix structure. The issue we encountered is related to the localizing matrices. The elements of these matrices usually contain polynomials -- provided that the original constraints were polynomials. This structure is easily described by the relaxation variables and the way an SDP is constructed, but Picos does not support summation of relaxation variables as required by the localizing matrices in Eq.(~\ref{localizing}). Instead, Picos introduces a quadratically growing number of equalities to define such elements, leading to unnecessarily large SDP formulations. We address these shortcomings to arrive at a fast implementation that generates a sparse SDP relaxation which also efficiently incorporates the polynomials in the entries of the localizing matrices.

Ncpol2sdpa has an implementation in Python. The source code for the implementation is available for download under GNU Public License\footnote{The package is available in the Python Package Index at \url{https://pypi.python.org/pypi/ncpol2sdpa/}}.

To develop a more general and scalable solution, we rely on an efficient library that support noncommuting algebras. SymPy is a Python module that has extensive support for noncommuting variables and has classes designed specifically for operator algebras~\cite{joyner2012open}.

The $F_{k}$ constraint matrices have a regular, block diagonal, sparse structure. The first block is diagonal, whereas the rest of the blocks are symmetric. The diagonal block encodes the symmetries in the moment matrix $M(y)$, but not the moment matrix itself, which is a symmetric block. The central data structure reproduce as much sparsity as possible in both implementations.

Ncpol2sdpa does not solve the SDP problem, it merely exports the relaxation to sparse SDPA format. SDPA is an efficient primal-dual interior point solver for SDPs that also has a distributed version~\cite{yamashita2003sdpara}, which is reportedly the fastest distributed solver~\cite{fujisawa2012high}.

\section{Toy example}\label{toyexample}
We provide a simple usage example here; this example comes with the code. More sophisticated applications are also supplied with the code.

Consider the following polynomial optimization problem of noncommuting variables~\cite{pironio2010convergent}:

\begin{eqnarray}
  \min_{x\in \mathbb{R}^2} & & x_1x_2 + x_2x_1\\
  \mathrm{s.t.} & & -x_2^2+x_2+0.5\geq 0\\
                & & x_1^2-x_1=0.
\end{eqnarray} 

Entering the objective function and the inequality constraint is easy. The equality constraint is a simple projection. We either substitute two inequalities to replace the equality, or treat the equality as a monomial substitution. The second option leads to a sparser SDP relaxation. The code samples below take this approach. In this case, the monomial basis is $\{1, x_1, x_2, x_1x_2, x_2x_1, x_2^2\}$. The corresponding relaxation is written as

\[ \min_{y}y_{12}+y_{21}\]

such that
\[
\left[\begin{array}{c|cc|ccc}
1 & y_{1} & y_{2} & y_{12} & y_{21} & y_{22}\\
\hline{}
y_{1} & y_{1} & y_{12} & y_{12} & y_{121} & y_{122}\\
y_{2} & y_{21} & y_{22} & y_{212} & y_{221} & y_{222}\\
\hline{}
y_{21} & y_{21} & y_{212} & y_{212} & y_{2121} & y_{2122} \\
y_{12} & y_{121} & y_{122} & y_{1212} & y_{1221} & y_{1222}\\
y_{22} & y_{221} & y_{222} & y_{2212} & y_{2221} & y_{2222}
\end{array} \right] \succeq{}0
\]

\[
\left[ \begin{array}{c|cc}
-y_{22}+y_{2}+0.5 & -y_{221}+y_{21}+0.5y_{1} & -y_{222}+y_{22}+0.5y_{2}\\
\hline{}
-y_{221}+y_{21}+0.5y_{1} & -y_{1221}+y_{121}+0.5y_{1} & -y_{1222}+y_{122}+0.5y_{12}\\
-y_{222}+y_{22}+0.5y_{2} & -y_{1222}+y_{122}+0.5y_{12} & -y_{2222}+y_{222}+0.5y_{22}
\end{array}\right]\succeq{}0.
\]
Apart from the matrices being symmetric, notice other regular patterns between the elements. These are taken care of as additional constraints in the implementation. The optimum for the objective function is $-3/4$. The Python implementation reads as follows:
\begin{verbatim}
from ncpol2sdpa import generate_variables, SdpRelaxation

# Number of Hermitian variables
n_vars = 2
# Order of relaxation
order = 2

# Get Hermitian variables
X = generate_variables(n_vars, hermitian=True)

# Define the objective function
obj = X[0] * X[1] + X[1] * X[0]

# Inequality constraints
inequalities = [-X[1] ** 2 + X[1] + 0.5]

# Equality constraints
equalities = []

# Simple monomial substitutions
monomial_substitution = {}
monomial_substitution[X[0] ** 2] = X[0]

# Obtain SDP relaxation
sdpRelaxation = SdpRelaxation(X)
sdpRelaxation.get_relaxation(obj, inequalities, equalities,
                             monomial_substitution, order)
sdpRelaxation.write_to_sdpa('example_noncommutative.dat-s')
\end{verbatim} 

This can be solved by invoking an SDP solver from the command line that takes a sparsely formatted SDPA problem as its input. Alternatively, if SDPA is available in the search path, we can obtain the primal and dual values by using a helper function in Ncpol2sdpa. This function writes the problem to a temporary file and solves it with SDPA. Continuing the example:

\begin{verbatim}
from ncpol2sdpa import solve_sdp
print(solve_sdp(sdpRelaxation))
\end{verbatim}
This should yield an output similar to this one:
\begin{verbatim}
(-0.7499997689892505, -0.7500000580502932)
\end{verbatim}

\section{Additional sparsity for binomial constraints}\label{additional}
Polynomial optimization problems of noncommuting variables often stem from applications in quantum physics. Constraints in such applications frequently take the form of commutators, anticommutators, orthogonality of operators, and idempotent operators. Take, for instance, the following optimization problem:
\[
\max_{E,\phi} \langle \phi, \sum_{ij} c_{ij} E_{i}E_{j}\phi \rangle
\]
subject to
\begin{eqnarray}
  ||\phi|| & = & 1 \nonumber \\
  E_{i}E_{j} & = & \delta_{ij}E_{i}\qquad \forall i,j \nonumber \\
  \sum_{i} E_{i} & = & 1 \nonumber \\
  \lbrack E_{i},E_{j}\rbrack & = & 0\qquad \forall i,j, \nonumber
\end{eqnarray}
where $\delta_{ij}$ stands for the Kronecker delta. The second constraint defines orthogonality and idempotency, whereas the last one is a commutator.

The na\"ive way of dealing with these equalities is to translate each to a pair of inequalities and generate the corresponding localizers (Section~\ref{generate}). This approach, however, will yield enormous relaxations, as the number of constraints scales quadratically with the number of variables.

An important observation is that such constraints consist of two simple monomials. Thus, the constraints of this type are simply binomials. Rearranging the terms of the binomials on both sides of the equations, we interpret these constraints as substitution rules. If an additional restriction is met so that the monomials should have a constant multiplier of either +1 or -1, then we use these substitution rules when generating the relaxation. Thus we eliminate the left-hand side of all occurring monomials, and replace them with the right-hand side. This leads to a repeated longest substring match problem of the monomials as the moment matrix and localizing matrices are generated. Similar substitutions are applied in the commutative case by GloptiPoly~3~\cite{henrion2009gloptipoly}.

While the substitution costs computational time, the length of the monomials on which the substitution is performed is short. For a second-order relaxation, the longest monomial will have four factors, making substitutions computationally feasible. The default substring matching algorithms of the symbolic libraries that we use are far more general than our special case: we provide faster heuristics to perform the replacements exploiting the simplicity of the matching and substitutions. 

The fast replacement heuristic pivots on the structure of monomials: there can only be a constant on the front, followed by a sequence of single variables or powers. In contrast, a generic symbolic formula can include arbitrary mathematical operation on the symbolic variables, including various unary, binary, and more complex function calls on the variables. By removing the need to check for the presence of such complexity, we already achieve a great speedup. We also rely on a form of lazy evaluation: we build a new return object only if a substring is matched, otherwise we return the original monomial. Since symbolic multiplication is a costly operation, this saves a tremendous amount of time. The overall improvement in running time is up to 10x in real-life problems. Cyclic substitutions are not detected and they will cause infinite loops. This is also true for the default replacement library. 

\section{Benchmarks}

\subsection{Benchmark optimization problem}\label{benchmarkexample}
We define the benchmark problem as follows:
\begin{eqnarray}
  \min_{H,X,\phi} & & \sum_{i,j}X_{i}X_{j} \\
  \mathrm{s.t.} & & X_{i}X_{j}=\begin{cases}
                                  1 & \mathrm{if} \quad i=j\\
                                  X_{j}X_{i} & \mathrm{otherwise}
                               \end{cases}
\end{eqnarray} 
The second type of equality translates to a monomial substitution, as described in Section~\ref{additional}.

We tested the scalability of Ncpol2sdpa by generating the relaxation of order one of this optimization problem on an increasing number of variables.

\subsection{Experimental settings}
We run the experiments on a workstation with an Intel Xeon E5620 processor and 24~Gbyte of main memory. The Python interpreter was version 2.6.8, the SymPy version was 0.7.2.

\begin{figure}[htb!]
  \begin{center}
      \includegraphics[width=0.8\columnwidth]{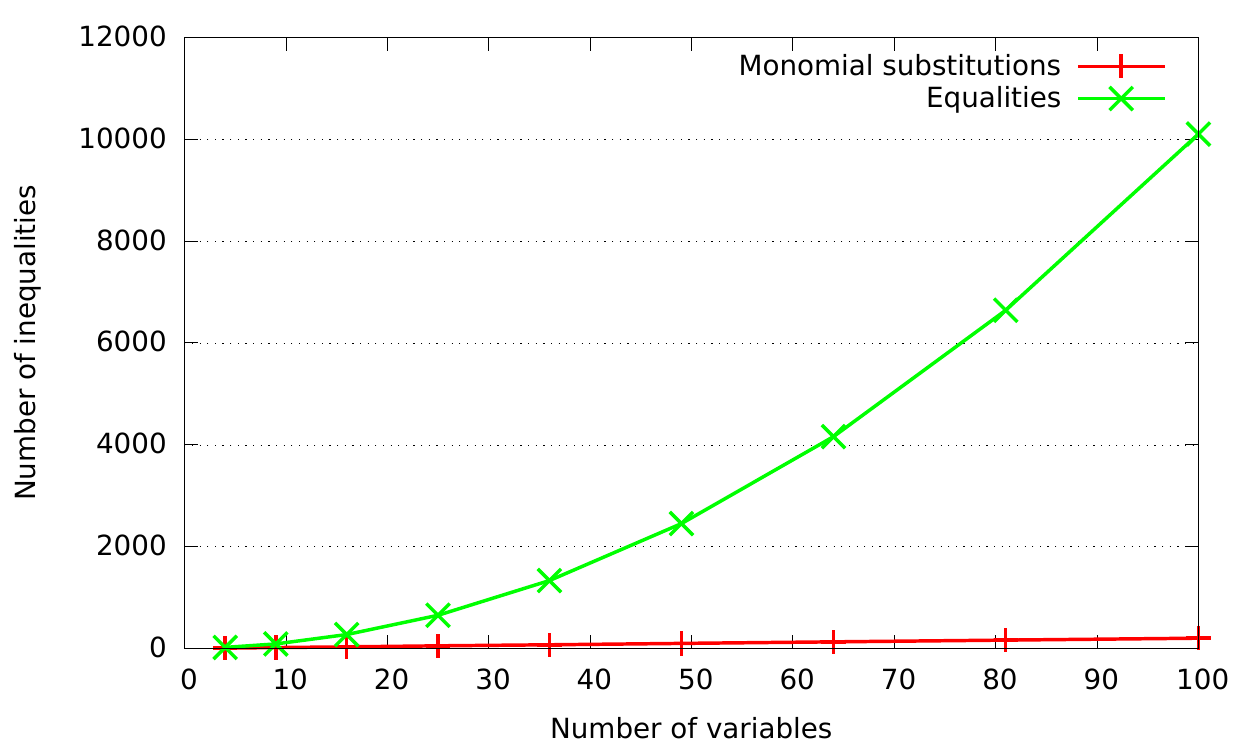}
    \caption{Comparison of the number of inequalities depending on how binomial constraints are handled. If we use monomial substitutions, the number of inequalities will grow linearly in the number of variables. Replacing all equalities with a pair of inequalities will lead to intractable SDP relaxations, as the quadratic growth of the number of inequalities shows.}
\label{nineq}
  \end{center}
\end{figure}

\begin{figure}[htb!]
  \begin{center}
      \includegraphics[width=0.8\columnwidth]{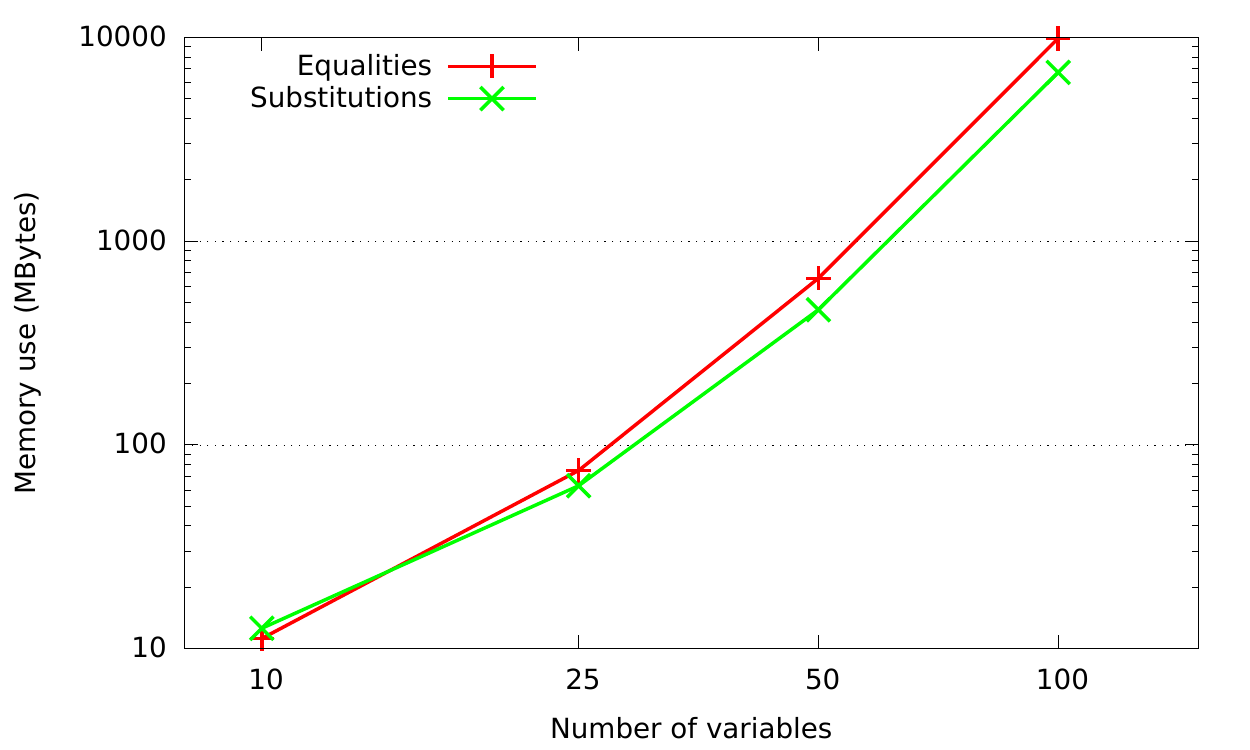}
    \caption{Memory use (log log scale). Equalities refer to case when equality constraints are exported as pairs of inequalities. Substitutions refers to the case where equalities are substituted as monomials.}
\label{memory}
  \end{center}
\end{figure}

Since the number of inequalities grew quadratically if we did not rely on monomial substitutions (Figure~\ref{nineq}), memory use was higher (Figure~\ref{memory}).

\subsection{Running time}
\begin{figure}[htb!]
  \begin{center}
      \includegraphics[width=0.8\columnwidth]{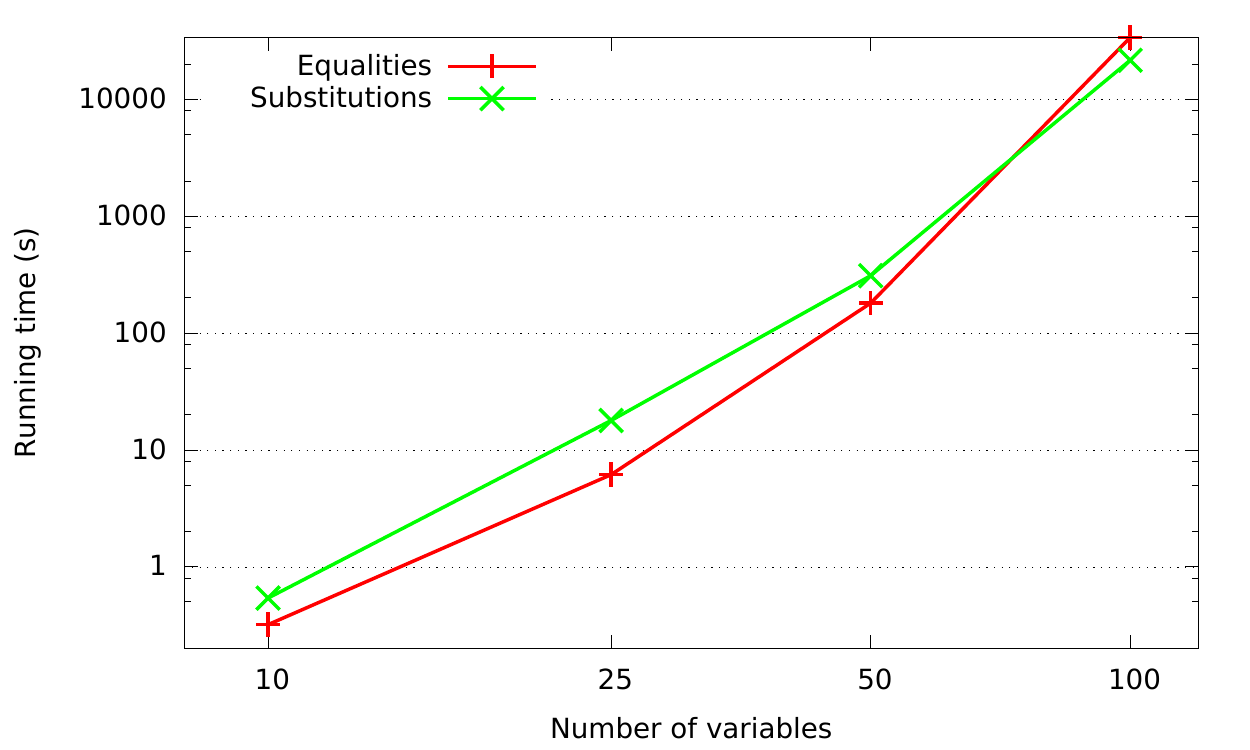}
    \caption{Running time (log log scale). Equalities refer to case when equality constraints are exported as pairs of inequalities. Substitutions refers to the case where equalities are substituted as monomials.}
\label{running}
  \end{center}
\end{figure}
If we used monomial substitutions, the running time was longer, even though we used our fast heuristic to perform the substitutions (Figure~\ref{running} and Table~\ref{tablerunning}). The trend, however, changes with a high number of variables. At this point, additional pairs of inequalities becomes more expensive to process then swapping the monomials on the fly.

\begin{table}
\begin{center}
\caption{Running time in seconds\label{tablerunning}}{
\begin{tabular}{l|lllll}
Variables & Equalities & Substitutions\\
\hline
10 & 0.32 & 0.54\\
25 & 6.15 & 17.87\\
50 & 180.3 & 310.36\\
100 & 33872.25 & 21632.12
\end{tabular}
}
\end{center}
\end{table}

\section{Related work}\label{relatedwork}
Polynomials of commutative variables have efficient supporting tools to generate the hierarchy of SDP relaxations of increasing order. The MATLAB toolbox Gloptipoly does this, and it is also capable of solving the SDPs by calling external libraries~\cite{henrion2003gloptipoly,henrion2009gloptipoly}. Showing the computational demand of the conversion, Gloptipoly is hardly able to deal with problems of a few dozen variables. This motivated the development of SparsePOP~\cite{waki2008algorithm}, which approximates the relaxation by extracting the correlative sparsity from the objective and the constrained polynomial. SparsePOP scales up to a thousand variables. However, if correlative sparsity cannot be identified, SparsePOP is limited to ten to thirty variables. We note in passing that Ncpol2sdpa natively supports commutative variables, hence it is an alternative to these tools if a license for MATLAB is not available.
 
\phantomsection
\label{par:comparisonwithothertools}
Tools for polynomials of noncommuting variables are harder to come by. The Mathematica package NCAlgebra can deal with such polynomials, but SDP relaxations are only addressed in unconstrained minimization problems~\cite{helton2012ncalgebra}. A similar toolbox, NCSOStools, exists for MATLAB~\cite{cafuta2011ncsostools} -- again, the key difference to our proposed solution is that NCSOStools only concerns unconstrained minimization problems. Yalmip is primarily a wrapper for various optimization problems in MATLAB~\cite{lofberg2004yalmip}, but it has an undocumented extension for noncommuting variables. Bermeja, a convex algebraic geometry package builds on this undocumented feature to solve NC problems~\cite{rostalski2012bermeja}. Unfortunately it does not scale beyond a few noncommuting variables, and in the most recent releases of Yalmip, the noncommuting variables are being phased out. Moreover, all of these tools require either Mathematica and MATLAB, which, apart from the cost of the license, presents complications if we would like to use high-performance cluster or cloud instances. Hence, our aim was to develop a scalable tool relying on free and open source software to generate the SDP hierarchy of relaxation for constrained polynomial optimization problems of noncommuting variables.

\section{Concluding remarks}
High-performance supporting libraries for solving SDPs do exist: they are technically capable of scaling to large problems. The missing link was a tool to generate the SDP relaxation of a given order from a polynomial optimization problem of noncommuting variables; this was developed and made available online under an open source license. The tool is able to generate the SDP relaxation of a hundred variables in about thirteen hours, although generating a sparser relaxation takes much longer. 

Improvements in the underlying symbolic library could improve execution time, especially by adding more efficient hashing functions and addressing the efficiency of substring replacement.

We further mention that a similar noncommuting optimization problem is discussed in \cite{cimpric2010method}, and for that formulation it is not possible to obtain a result with Ncpol2sdpa. The SDP relaxation in \cite{pironio2010convergent} converges to the solution of a double optimization: one over the operators, and another one on the vectors of norm one. This latter restriction is not present in the optimization of \cite{cimpric2010method}. Solving this more generic problem remains for future work.

\section{Acknowledgement}
This work was supported by the European Commission Seventh Framework Programme under Grant Agreement Number FP7-601138 PERICLES, by the AWS in Education Machine Learning Grant award, and by the Red Espa\~nola de Supercomputaci\'on grants number FI-2013-1-0008 and FI-2013-3-0004. 

\bibliographystyle{apalike}
\bibliography{bibliography}

\begin{thebibliography}{}

\bibitem[Cafuta et~al., 2011]{cafuta2011ncsostools}
Cafuta, K., Klep, I., and Povh, J. (2011).
\newblock {NCSOStools}: a computer algebra system for symbolic and numerical
  computation with noncommutative polynomials.
\newblock {\em Optimization Methods and Software}, 26(3):363--380.

\bibitem[Cimpri\v{c}, 2010]{cimpric2010method}
Cimpri\v{c}, J. (2010).
\newblock A method for computing lowest eigenvalues of symmetric polynomial
  differential operators by semidefinite programming.
\newblock {\em Journal of Mathematical Analysis and Applications},
  369(2):443--452.

\bibitem[Fujisawa et~al., 2012]{fujisawa2012high}
Fujisawa, K., Sato, H., Matsuoka, S., Endo, T., Yamashita, M., and Nakata, M.
  (2012).
\newblock High-performance general solver for extremely large-scale
  semidefinite programming problems.
\newblock In {\em Proceedings of SC-12, International Conference on High
  Performance Computing, Networking, Storage and Analysis}, pages 93:1--93:11.

\bibitem[Helton et~al., 2012]{helton2012ncalgebra}
Helton, J., Miller, R., and Stankus, M. (2012).
\newblock {NCAlgebra}: a {Mathematica} package for doing non-commuting algebra.

\bibitem[Henrion and Lasserre, 2003]{henrion2003gloptipoly}
Henrion, D. and Lasserre, J. (2003).
\newblock {GloptiPoly}: Global optimization over polynomials with {Matlab} and
  {SeDuMi}.
\newblock {\em ACM Transactions on Mathematical Software}, 29(2):165--194.

\bibitem[Henrion et~al., 2009]{henrion2009gloptipoly}
Henrion, D., Lasserre, J., and L{\"o}fberg, J. (2009).
\newblock {GloptiPoly} 3: moments, optimization and semidefinite programming.
\newblock {\em Optimization Methods \& Software}, 24(4-5):761--779.

\bibitem[Joyner et~al., 2012]{joyner2012open}
Joyner, D., {\v{C}}ert{\'\i}k, O., Meurer, A., and Granger, B.~E. (2012).
\newblock Open source computer algebra systems: {SymPy}.
\newblock {\em ACM Communications in Computer Algebra}, 45(3/4):225--234.

\bibitem[Lasserre, 2001]{lasserre2001global}
Lasserre, J. (2001).
\newblock Global optimization with polynomials and the problem of moments.
\newblock {\em SIAM Journal on Optimization}, 11(3):796--817.

\bibitem[L\"ofberg, 2004]{lofberg2004yalmip}
L\"ofberg, J. (2004).
\newblock {YALMIP}: A toolbox for modeling and optimization in {MATLAB}.
\newblock In {\em Proceedings of CACSD-04, IEEE International Symposium on
  Computer Aided Control Systems Design}, pages 284--289.

\bibitem[Navascu\'es et~al., 2013]{navascues2013paradox}
Navascu\'es, M., Garc\'ia-S\'aez, A., Ac\'in, A., Pironio, S., and Plenio, M.
  (2013).
\newblock A paradox in bosonic energy computations via semidefinite programming
  relaxations.
\newblock {\em New Journal of Physics}, 15(2):023026.

\bibitem[Navascu{\'e}s et~al., 2008]{navascues2008convergent}
Navascu{\'e}s, M., Pironio, S., and Ac{\'i}n, A. (2008).
\newblock A convergent hierarchy of semidefinite programs characterizing the
  set of quantum correlations.
\newblock {\em New Journal of Physics}, 10(7):073013.

\bibitem[Navascu{\'e}s et~al., 2012]{navascues2012sdp}
Navascu{\'e}s, M., Pironio, S., and Ac{\'\i}n, A. (2012).
\newblock {\em Handbook on Semidefinite, Conic and Polynomial Optimization},
  chapter {SDP} Relaxations for Non-Commutative Polynomial Optimization, pages
  601--634.
\newblock Springer.

\bibitem[Pironio et~al., 2010]{pironio2010convergent}
Pironio, S., Navascu\'es, M., and Ac\'in, A. (2010).
\newblock Convergent relaxations of polynomial optimization problems with
  noncommuting variables.
\newblock {\em SIAM Journal on Optimization}, 20(5):2157--2180.

\bibitem[Rostalski, 2012]{rostalski2012bermeja}
Rostalski, P. (2012).
\newblock \url{http://math.berkeley.edu/~philipp/Main/HomePage}.

\bibitem[Sagnol, 2012]{sagnol2012picos}
Sagnol, G. (2012).
\newblock Picos documentation.
\newblock Technical Report 12-48, Zuse Institut Berlin.

\bibitem[Vandenberghe and Boyd, 1996]{vandenberghe1996semidefinite}
Vandenberghe, L. and Boyd, S. (1996).
\newblock Semidefinite programming.
\newblock {\em SIAM Review}, 38(1):49--95.

\bibitem[Waki et~al., 2008]{waki2008algorithm}
Waki, H., Kim, S., Kojima, M., Muramatsu, M., and Sugimoto, H. (2008).
\newblock Algorithm 883: Sparsepop---a sparse semidefinite programming
  relaxation of polynomial optimization problems.
\newblock {\em ACM Transactions on Mathematical Software}, 35(2):15.

\bibitem[Yamashita et~al., 2003]{yamashita2003sdpara}
Yamashita, M., Fujisawa, K., and Kojima, M. (2003).
\newblock {SDPARA}: Semidefinite programming algorithm parallel version.
\newblock {\em Parallel Computing}, 29(8):1053--1067.

\end{thebibliography}

\end{document}